\documentstyle[aps,prl,epsf,epsfig,12pt]{revtex}

\newcommand\simlt{\lower.5ex\hbox{$\; \buildrel < \over \sim \;$}}
\newcommand\simgt{\lower.5ex\hbox{$\; \buildrel > \over \sim \;$}}

\title{Baryon-poor outflows from rotating black holes}
\author{Maurice H.P.M. van Putten}

\begin{document}
\maketitle
\baselineskip15pt
\begin{abstract}  
\baselineskip15pt
\mbox{}\\
Cosmological gamma-ray bursts are probably powered by systems harboring
a rotating black hole. We show that frame-dragging creates baryon-poor 
outflows in a differentially rotating gap along an open magnetic flux-tube
with dissipation ${P}=\Omega_T(\Omega_H-2\Omega_T)A_\phi^2$, 
where $2\pi A_\phi$ denotes the flux in the open tube, for angular 
velocities $\Omega_H$ and $\Omega_T$ of the black hole and the torus,
respectively. The output in photon-pair ouflow will be reprocessed
in an intervening hypernova progenitor wind or the ISM.

\mbox{}\\
{\em Subject headings}: gamma-rays: bursts, theory
\mbox{}\\
\end{abstract}

Cosmological gamma-ray bursts may be powered by rotating black holes
following the collapse of young massive stars\cite{woo93} in
hypernovae \cite{pac98,bro00} or the coalescence of black hole-neutron 
star binaries \cite{pac91}. If all black holes are produced by stellar 
collapse, they should be nearly maximally rotating \cite{bar70,bet98}, 
whose mass is relatively high \cite{bro99,mvp01}. The torus is expected 
to form from fallback matter stalled against an angular momentum barrier 
or as the debris of a neutron star following tidal break-up around a 
Kerr black hole. The torus will provide a similarly-shaped magnetosphere 
from the remnant magnetic field. The magnetic field strength derives 
from conservation of magnetic flux and linear amplification
(see \cite{pac98,klu98}).

GRB/afterglow studies indicate an output in ultrarelativistic baryon 
poor jets. In this {\it Letter}, we show that frame-dragging creates powerful
photon-pair outflows in a differentially rotating gap along open magnetic flux-tubes
supported by black holes in equilibrium with a surrounding torus magnetosphere.
This process is continuous and involves no baryonic load. This analysis represents 
a non-perturbative extension of earlier calculations on pair-production about the
a Wald field \cite{wal74,mvp00,hey01}.

Pair-creation in magnetospheres around rotating black holes is made possible by
the Rayleigh criterion, since radiation from the horizon posesses a specific
angular momentum at least twice that of the black hole itself. The black hole 
may support open flux-tubes by an magnetic moment in equilibrium with a surrounding 
torus magnetosphere. This is a consequence of a no fourth-hair theorem (see below). 
Frame-dragging introduces Faraday potential drops across these flux-tubes. This
creates a competition between local equilibration into a dissipation-free state
and current-continuity as a global constraint. Here, currents are determined by
slip-slip boundary conditions on the horizon and infinity.  As will be shown, the 
outcome of this competition is dissipation in a differentially rotating gap. This
results in the creation of a photon-pair outflow to infinity.

A torus surrounding a black hole faces the black hole horizon on the inside and 
infinity on the outside. These ``two faces" are each equivalent in poloidal 
topology to pulsar magnetospheres with commensurate causal interactions, as
shown in Fig. 1.
The inner face of the torus receives energy and angular momentum, as does
a pulsar when infinity wraps around it (see Fig. 2 of \cite{mvp99b}); 
the outer face always looses energy and angular momentum similar to
ordinary pulsars \cite{gol69}. 
Here, we assign Dirichlet boundary conditions on field-lines footed in the torus, 
and outgoing/ingoing radiative boundary conditions at infinity/on the horizon. 
The former represent no-slip boundary conditions in the non-perturbative limit of 
flux-tubes in equilibrium charge-separation, whereas the latter represent slip 
boundary conditions in the same limit. It will be appreciated that field-lines with 
either $DR-$ or $D\bar{R}-$boundary conditions correspond to open field-lines in 
pulsar magnetospheres: those which pass through the light cylinder (lc) out to 
infinity \cite{gol69}. In the equilibrium charge-separation limit, the inner and 
outer torus magnetospheres rotate with the approximately Keplerian angular velocity 
of the torus. The inner torus magnetosphere passes through an inner light surface 
(ils) \cite{zna77} close to the horizon, as the limiting surface where particles 
could stay in closed orbits with the angular velocity $\Omega_T$ of the torus. The 
ils and ols serve the same role as the outer light cylinder in a pulsar magnetosphere. 
This equivalence to pulsars indicates that the torus mediates two Poynting-flux 
dominated winds, one flowing into the black hole and one flowing out to infinity - 
each bringing about energy and angular momentum transport. If the black hole spins 
at the rate of the torus times a geometrical factor, either without \cite{mvp01} or 
with \cite{mvp01d} gravitational radiation, a state of suspended accretion may result
in which the net torque vanishes. The suspended accretion state hereby lasts for the 
life-time of rapid spin of the black hole.

In the lowest energy state, a rotating black hole surrounded by a torus
magnetosphere assumes a magnetic moment $\mu_H$. This is no fourth-hair. 
Indeed, we have \cite{car68,coh73,wal74}
\begin{eqnarray}
\mu_H=qJ_H/M
\label{EQN_MM}
\end{eqnarray}
with induced horizon flux $\phi_H=4\pi qM\Omega_H$ \cite{wal74,dok87,mvp00}. This may 
be understood to form an azimuthal current $q\Omega_H/2\pi$ generated by corotation 
of a horizon charge charge $q$. Exposed to an external uniform magnetic field aligned 
with its axis of rotation, the potential energy of the electromagnetic field is 
\begin{eqnarray}
{\cal E}\approx q^2/2r_H-\mu_H B.
\label{EQN_EE}
\end{eqnarray}
The ground state assumes $\partial{\cal E}/\partial q=0$, which defines 
$q^e\approx BJ_H(r_H/M)$ in view of (\ref{EQN_MM}).
(It can be shown that $\delta (J_H/M)/\delta q=O(B^2M^2)$, which is
neglible for gravitationally weak magnetic fields.)
This equilibrium charge may develop by accretion \cite{wal74} or by pair-creation 
\cite{gib75}. Combined with the horizon flux in the uncharged state, the net flux 
becomes
\begin{eqnarray}
\Phi_H^e=4\pi BM^2\cos\lambda+\phi_H^e\approx 4\pi BM^2,
\label{EQN_QH}
\end{eqnarray}
where $\sin\lambda=a/M$. Here, the numerical value on the right hand-side is given 
for Wald's value $q^e=2BJ_H$, which results from an exact analysis in a source-free 
Wald-field. The energy arguments used describe black hole coupling to the
poloidal magnetic field average $B=<B_z>$ over an $O(M)$-neighborhood. 
The result is therefore expected to be robust within a factor of unity for 
general, aligned magnetospheres, and is in agreement with an explicit calculation in 
the force-free limit \cite{lee01}. In particular, it should not depend too 
sensitively on the behavior of magnetic winds as they enter the horizon.
The magnetic moment in the magnetostatic ground state satisfies 
$\mu_H^e\approx 2BMr_H^2(2M\Omega_H)^2$ which, by $\phi_H$, produces an azimuthal 
component of the electromagnetic vector potential
\begin{eqnarray}
A_{\phi}^{(\mu)}=\frac{\mu_H^e}{r_H}.
\label{EQN_A0}
\end{eqnarray}
The magnetic moment in the ground state hereby serves {\em two} purposes,
particularly when the black hole rotates rapidly: it ensures strong black hole-torus 
coupling and it permits the horizon to support an open flux-tube infinity.
The open flux-tube is endowed with conjugate radiative-radiative
$R\bar{R}-$boundary conditions, which become slip-slip boundary conditions in
the equilibrium charge-separation limit (below).  
Considerable disorder (e.g.: in spectral energy-density)
is expected in the external magnetic field due to turbulent shear within
the torus \cite{mvp99b}. In contrast, the open field-lines along the axis of rotation
are as always well-ordered, since they are supported by $\mu^e_H$ via (\ref{EQN_A0}).

Open flux-tubes can be created out of closed loops between the black hole and the 
torus, as shown in Fig. 2. This change in topology may be compared with coronal mass 
ejections \cite{mvp01e}. Thus, the magnetic flux of the open flux-tube supported by 
$\mu_H^e$ is equal and opposite to the magnetic flux of a surrounding flux tube 
supported by the surrounding torus. With reference to the the boundary conditions, 
we form $D\bar{R}\rightarrow DR+ R\bar{R}$ with fractions of magnetic flux
\begin{eqnarray}
f_{DR}=f_{R\bar{R}}=f_o,
\label{EQN_FO}
\end{eqnarray}
e.g., relative to the net poloidal flux $2\pi A$ supported by the torus.
This transformation is due, in part, to the flow of baryonic
matter into the equatorial plane where it settles into a torus 
and, in part, due to bulging of the inner torus magnetosphere
in response to repulsive current-current interactions in the poloidal
current loop - a fast magnetosonic wave as a nonlinear feature to the powerful 
DC Aflv\'en wave in the torus magnetosphere around the black hole.

Locally, the magnetosphere tends to equilibrate into a non-dissipative 
state of equilibrium charge-separation. In particular, this will hold true for 
the inner and outer torus magnetospheres, by their equivalence to pulsar 
magnetospheres.
Recall that a magnetosphere surrounding a pulsar with angular velocity $\Omega_{psr}$ 
assumes an equilibrium charge-density at the Goldreich-Julian value 
$\rho=-\Omega_{psr}B/2\pi$ on-axis, where $B$ denotes the 
poloidal magnetic field-strength \cite{gol69}. Note that the near-axis field-lines in
pulsar magnetospheres are endowed with $DR-$boundary conditions. In general, 
equilibrium flux surfaces are described by two parameters: an angular velocity and a 
poloidal current (see \cite{tho86}). It becomes of interest to consider the 
competition between these local equilibrium considerations and current 
continuity as a global constraint.

 Locally, equilibrium flux-surfaces $A_\phi$ assumes a uniform electrostatic 
 potential $-A_t=\Omega A_\phi$ with rigid angular velocity $\Omega$ in 
 Boyer-Lindquist coordinates (see \cite{tho86}). The associated equilibrium 
 charge-density $\rho^e$ can be calculated from Maxwell's equations. About the 
 axis of rotation, these give
\begin{eqnarray}
\left(\sqrt{-g}g^{ac}g^{td}F_{cd}\right)_{,a}=-4\pi\sqrt{-g}j^t,
\end{eqnarray}
where $\sqrt{-g}=\rho^2\sin\theta$, and, by evaluation in Boyer-Lindquist 
coordinates, 
\begin{eqnarray}
 [\sin\theta(\xi^aA_a)_{,\theta}]_{,\theta} \simeq 4\pi\sqrt{-g}\alpha^2 j^t
\end{eqnarray}
about the axis of rotation $\theta=0$. The equilibrium charge-density 
$\rho^e=\alpha^2 j^t$ (one $\alpha$ for redshift and one for volume density) as 
seen by zero-angular momentum observers (ZAMOs) assumes the asymptotic values 
\begin{eqnarray}
\rho^{e}=-(\Omega+\beta)B/2\pi=\left\{
\begin{array}{lr}
(\Omega_H-\Omega_+)B/2\pi & \mbox{on~}H,\\
-\Omega_- B/2\pi & \mbox{at~}\infty,
\end{array}
\right.
\label{EQN_RHO}
\end{eqnarray} 
where $\Omega_+$ and $\Omega_-$ denote the angular velocities of the
equilibrium sections attached to the horizon and infinity, respectively. This 
on-axis density distribution is distinct from the Goldreich-Julian charge-density 
in pulsar magnetospheres most notably so in a sign-change, whenever 
$\Omega_H-\Omega_+>0$ and $\Omega_->0$. Indeed, when $\Omega_H-\Omega>0$, the 
charge-density (\ref{EQN_RHO}) possesses an interface with 
$\rho^e=-(\Omega+\beta)B/2\pi =0$ \cite{hir98}. This defines an underlying 
$pn-$structure to the flux-tube. With $B$ parallel to $\Omega_H$, the $n-$section 
faces the horizon and the $p-$section faces infinity. The $pn-$structure permits 
the inner flux-cone to conduct a DC current towards the black hole, given 
a lower electrostatic potential of the $n-$section.

The current in the asymptotically ultrarelativistic flows is convection 
of the equilibrium charge-density (\ref{EQN_RHO}). Thus, $j^2\rightarrow0$ 
describes the continuum limit of the $R\bar{R}-$boundary conditions, wherein
drift currents are suppressed by the high Lorentz factor bulk flow upon going
into the horizon \cite{pun90} as well as going out to infinity \cite{mvp01c}.
This defines two current-sources in series (Fig. 3). Subject to current continuity, 
we have:
\begin{eqnarray}
-I=(\Omega_H-\Omega_+)A_\phi=\Omega_-A_\phi.
\end{eqnarray}
Here, $I<0$ refers to currents towards the black hole. Thus, $I$ generally arises 
from differential rotation between the $+$ and $-$ sections. The commensurate
Faraday-induced potential difference
\begin{eqnarray}
 \Delta V=[\Omega]A_\phi =(\Omega_+-\Omega_-)A_\phi=\Omega_HA_\phi+2I,
\label{EQN_F1}
\end{eqnarray}
across a dissipative inversion layer in differential rotation.
Likewise, the asymptotic state of the outer flux-cone endowed with
$DR-$boundary conditions defines a current-source at infinity.
Current closure at infinity and (\ref{EQN_FO}) give
$I=\Omega_TA_\phi$ and hence $\Omega_-=\Omega_T$. 
It follows that the inversion layer dissipates energy at a rate \cite{mvp01c}
\begin{eqnarray}
P=I\Delta V=\Omega_T(\Omega_H-2\Omega_T) A_\phi^2
\label{EQN_L1}
\end{eqnarray}
provided that $\Omega_H>2\Omega_T$. This shows that the open 
field-lines along the axis of rotation contain a region which
is out of equilibrium as a consequence of current continuity. This
manifests itself in a macroscopic photon-pair outflow. Note that
this involves no baryonic load.

The sign change in the charge density in the gap introduces some common
ground with outer gaps in pulsar magnetospheres \cite{hir00}, though the
boundary conditions are different.
The equilibrium charge $\rho_e$ is approximately linear in the inversion layer 
about the root $r=r^*$ of $\Omega+\beta^*=0$. Here, $\Omega=(\Omega_++\Omega_-)/2$ 
and so
\begin{eqnarray}
\rho_e(z)\approx -bz,~~z=r-r^*,
\label{EQN_ODD}
\end{eqnarray}
where $b=\beta_{,r}^*$, being positive in the $n-$section
and negative in the $p-$secion. In the ultrarelativistic limit, the current density
$j=-n$ in the present sign convention, where $n=n_++n_-$ denotes the sum of the 
charge densities $e^+$ and $e^-$. In the stationary limit, $j$ is constant as a 
function of height upon ignoring curvature drift. Since $n_\pm$ derive from 
dissipation in the gap, the associated local charge density satisfies 
\begin{eqnarray}
\rho_j=j\frac{[V]_{-h/2}^z-[V]_z^{h/2}}{[V]_{-h/2}^{h/2}},
\label{EQN_RJ}
\end{eqnarray}
where $E=-V^\prime(z)$ is the electric field in terms of the
electrostatic potential $V$. $\Delta V$ shall denote
$[V]_{-h/2}^{h/2}=[\Omega]A_\phi\le [\beta]^{h/2}_{h/2}A_\phi$.
Thus, $E$ satisfies Poisson's equation $E^\prime=4\pi(\rho+\rho_j),$ 
where $\rho(z)=-\rho_{e}(z).$ The equations (\ref{EQN_ODD}) and (\ref{EQN_RJ}) give
\begin{eqnarray}
\rho_j\sim 2jz/h~~~(-h/2<z<h/2).
\end{eqnarray}
No new net charges are created and, in the linear regime (\ref{EQN_ODD}), the net 
charge within the inversion layer remains zero. Hence, the electric field is non-zero
only within the layer, leaving zero surface-charge density 
on the two virtual Faraday disks at $z=\pm h/2$. For the front at 
$z=h/2$, the outgoing radiation
pressure $P_{r}$ derives from from particles moving upwards, i.e.: $n_-=-(\rho+j)/2$ 
in the present sign convention. Hence, we arrive at an output
\begin{eqnarray}
L_{p}=\int_{-h/2}^{h/2}n_-Edz=-\frac{1}{2}j\Delta V=\frac{1}{2}
\Omega_T(\Omega_H-2\Omega_T)A_\phi^2,
\label{EQN_L2}
\end{eqnarray}
using (\ref{EQN_L1}).
Thus, one-half the power converted in the inversion layer
produces $e^\pm\gamma$-outflow. The output power
in isotropic equivalent luminosity assumes GRB values for
canonical values of $B=10^{15}$G in the torus magnetosphere. The Lorentz factor of
the outflow will be high, whose terminal value depends, in part, on curvature 
radiation in the open field-lines. At the distance where the GRB takes place, 
it further depends on the amount of baryonic matter intercepted from the
interstellar medium, as described in current theories of GRB-afterglows.

The dissipation (\ref{EQN_L1}) results a photon-pair outflow (\ref{EQN_L2}) to 
infinity. Observed emissions result from reprocessing in the intervening medium,
notably a hypernova progenitor wind or the interstellar medium
\cite{li01}.

The BATSE GRB catalogue shows a bi-modal distribution of short durations of about 
0.3s and long durations of about 30s \cite{kou93,pac99}. 
In \cite{mvp01}, we associate this bi-modal distribution with hyper-/suspended
accretion states onto slowly/rapidly spinning black holes.
For black hole angular velocities $\Omega_H>2\Omega_T$ while sufficiently slow to 
assume a state of hyperaccretion, baryon poor outflows will arise similarly as
in the suspended accretion state. Thus, we predict that HETE-II could detect 
afterglows also from short bursts, but with weak or no iron-line emissions 
\cite{mvp01}.
A recent analysis \cite{fra01} of the opening angle from long bursts, inferred from 
a break in the temporal evolution of the GRB luminosity, suggests a narrow 
distribution of GRB fluence around $5\times 10^{50}$ergs with a considerable
spread in the half-opening angles $\theta_j$. In \cite{mvp01b}, we attribute this
to a standard half-opening angle $\theta_H\simeq 35^{\mbox{o}}$ of the open flux-tube
on the horizon in the presence of a geometrically thick torus. In this scenario,
the spread in the observed half-opening angles $\theta_j$ on the celestial sphere 
is due variations in the luminosity of the collimating wind \cite{lev00}.
This may be tested in future observations, wherein the estimated value of $\theta_H$ 
should provide a cut-off to $\theta_j$.

\mbox{}\\
{\bf Acknowledgements.}
This work is partially supported by NASA Grant 5-7012, the MIT C. Reed Fund and a 
NATO Collaborative Linkage Grant. The author thanks the hospitality of KIAS,
Caltech, SUNY at Stony Brook and JINR, Dubna, where some of this work was performed, 
and gratefully acknowledges stimulating  discussions with G.E. Brown, P. Goldreich,
E.C. Ostriker, A. Levinson and the referees for constructive comments.

\newpage
\centerline{{\bf Figure captions}}
\mbox{}\\
\mbox{}\\
{\bf Figure 1.} {\small
Equivalence in poloidal topology of the magnetosphere of torus to that of a pulsar 
indicates commensurate causal interactions in a black hole-torus system in suspended 
accretion. The torus magnetosphere is supported by a magnetic moment density $\mu_T$ 
in the surrounding magnetized matter. In equilibrium, the black hole assumes a
a magnetic moment $\mu_H^e$ which supports a tube of open field-lines along its axis 
of rotation. These open field-lines have conjugate radiative-radiative 
$R\bar{R}-$boundary conditions or, in the equilibrium charge-separated limit,
slip-slip boundary conditions on the horizon and infinity. These field-lines have
no counterpart in pulsar magnetospheres. Field-lines from the
torus extending to the black hole have $D\bar{R}-$boundary conditions, and those
extending to infinity have $DR-$boundary conditions. These are equivalent to open
field-lines in pulsar magnetosperes. A limited number of field-lines
make up closed field lines in an inner and outer bag attached to the two
faces of the torus with $DD-$boundary conditions, delineated
by the dashed curces representing the inner light surface
and the outer light cylinder.
This equivalence shows that the inner face of the torus receives energy and
angular momentum as does a pulsar when infinity wraps around it \cite{mvp01e},
while the outer face always looses energy and angular momentum as do ordinary
pulsars. The strength of these interactions is given by the equivalent angular 
velocities $-\Omega_{psr}=\Omega_H-\Omega_T$ and $\Omega_{psr}=\Omega_T$, where 
$\Omega_H$ and $\Omega_T$ are the angular velocities of the black hole and the 
torus. (Reprinted from \cite{mvp01c}, \copyright2001, Elsevier B.V.)
}

\mbox{}\\
{\bf Figure 2.} {\small 
Topological diagram of the creation of a co-axial flux-cone in a black hole-torus 
system. The torus supports a similarly shaped magnetosphere, which comprises upper
and lower annular tubes connected to the black hole. Bulging of the these tubes in
a poloidal expansion of their upper and outer layers is induced by repulsive
current-current interactions -- a DC fast magnetosonic wave as a nonlinear feature
to strong Alfv\'en waves associated with the powerful interaction between the 
black hole and the torus. A pair of co-axial flux-tubes forms following a 
stretch-fold-cut on the upper annular tube with end points taken to infinity 
($\alpha$). This leaves the inner and outer tubes with magnetic flux equal in
magnetic and opposite in sign ($\beta$). (Reprinted from \cite{mvp01c},
\copyright2001, Elsevier B.V.)
}

\mbox{}\\
{\bf Figure 3.} {\small 
The open flux-tube along the axis of rotation is endowed with slip-slip
boundary conditions on the horizon and infinity, described by angular
velocities $\Omega_+$ and $\Omega_-$, respectively. 
The angular velocities as observed by local zero-angular momentum
observers define the equilibrium charge-separation densities (\ref{EQN_RHO}).
Ultrarelativistic flows introduce $j^2\rightarrow0$ on the horizon
\cite{pun90} and to infinity, which introduces current sources
$I_+=(\Omega_H-\Omega_+)A_\phi$ and $I_-=\Omega_-A_\phi$, where $2\pi A_\phi$ 
denotes the flux in the flux-tube. Current continuity imposes the constraint 
$I_+=I_-$. Differential rotation $\Omega_+\ne\Omega_-$ hereby gives rise to 
a Faraday-induced potential drop between these lower and upper sections.
}
\newpage
\begin{center}
\mbox{}\\
\vskip1in
\epsfig{file=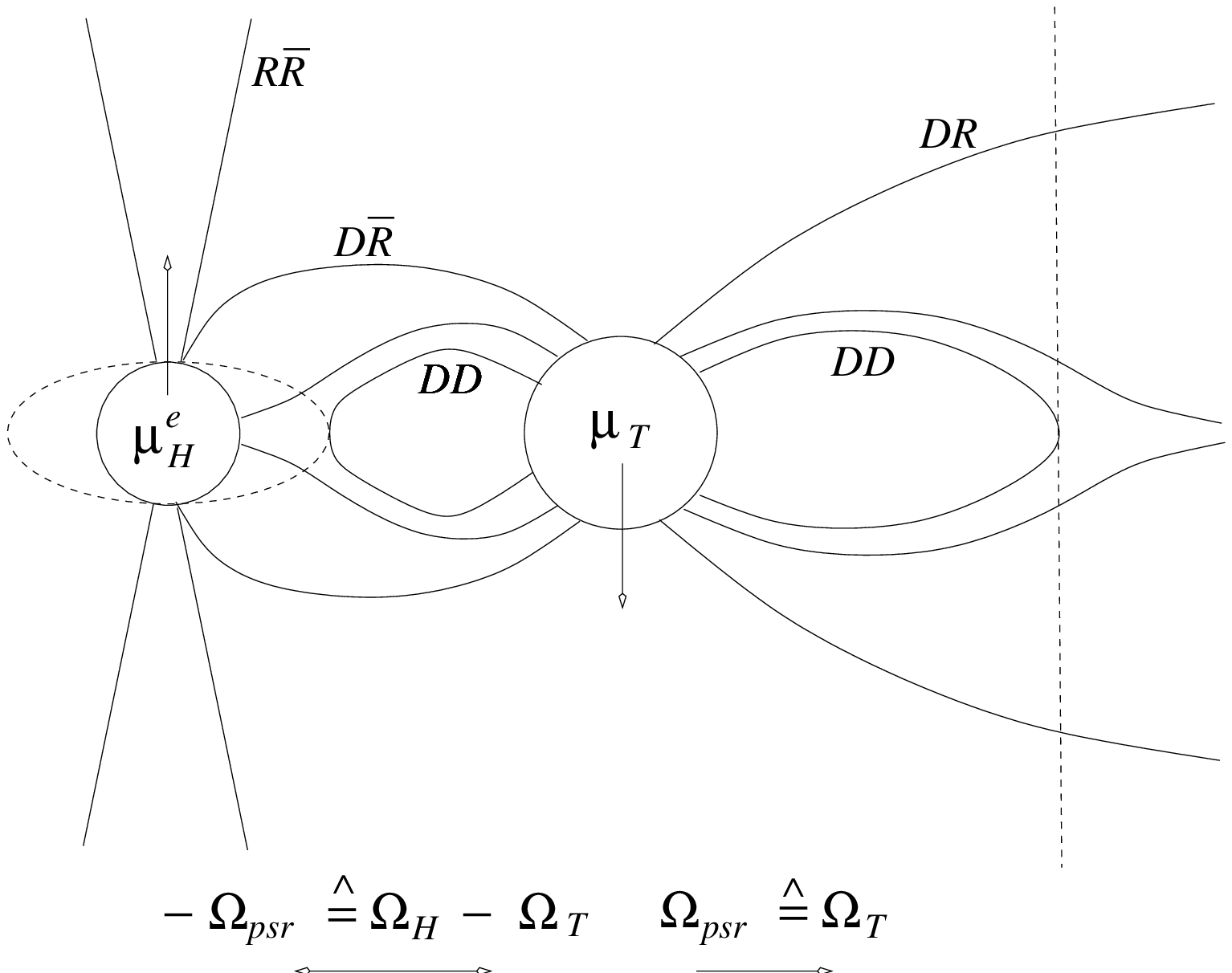}
\end{center}
\vskip2in
{\sc FIGURE 1}
\newpage
\begin{center}
\mbox{}\\
\vskip1in
\epsfig{file=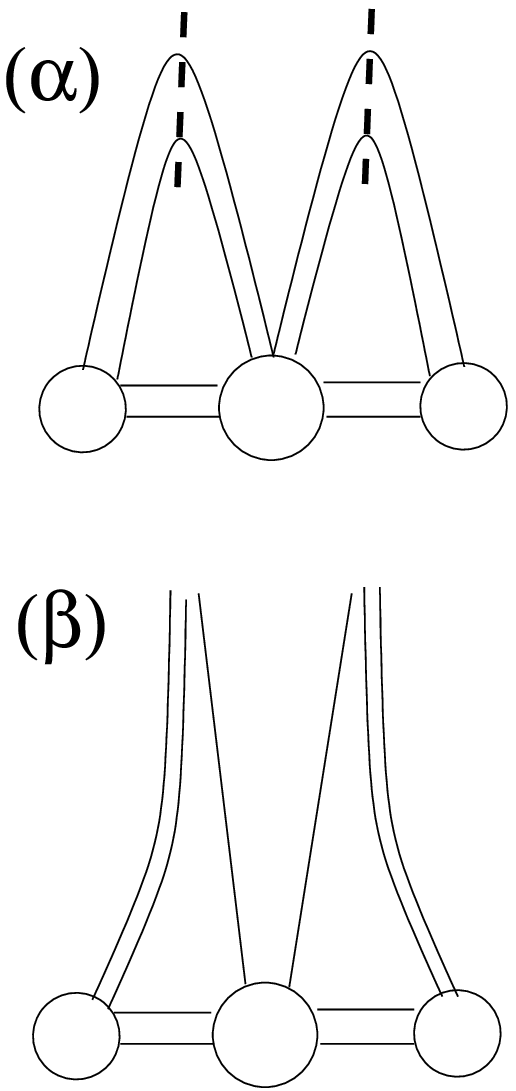}
\end{center}
\vskip2in
{\sc FIGURE 2}
\newpage
\begin{center}
\mbox{}\\
\vskip1in
\epsfig{file=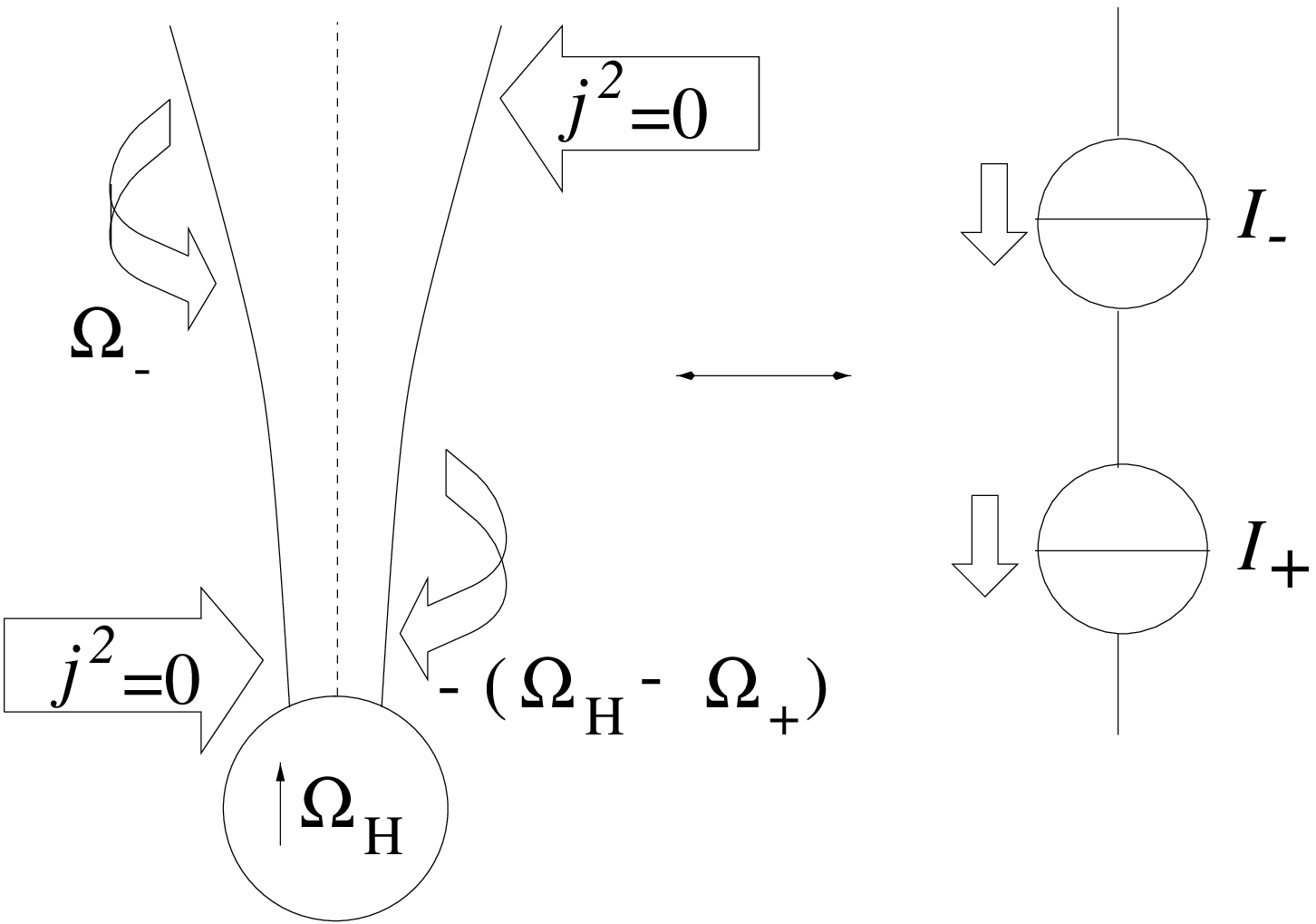}
\vskip2in
\end{center}
{\sc FIGURE 3}
\end{document}